\documentclass[10pt,a4paper]{article} 

\usepackage{anysize}
\usepackage[format = hang]{caption}
\usepackage{amsmath,amsthm,verbatim,amssymb,amsfonts,amscd, graphicx}
\usepackage{graphics,natbib}
\defcitealias{Lid1942}{Nils Lid}

\usepackage{titlesec,footmisc}
\usepackage{hyperref} 

\usepackage{listings}
\usepackage{booktabs} 

\theoremstyle{plain}

\theoremstyle{definition}

\def\jilek{{J\'{\i}lek}}

\def\E{{\rm E}}

\def\hatt{\widehat}
\def\arr{\rightarrow}
\def\N{{\rm N}}

\def\beq{\begin{eqnarray}}
\def\eeq{\end{eqnarray}}

\def\beqn{\begin{eqnarray*}}  
\def\eeqn{\end{eqnarray*}}

\def\E{{\rm E}}

\def\N{{\rm N}}

\def\Pr{P}

\def\quadandquad{\quad {\rm and} \quad}
\def\arr{\rightarrow}
\def\hatt{\widehat}

\def\sumin{\sum_{i=1}^n}

\def\rootn{\sqrt{n}}

\def\prof{{\rm prof}}

\def\simm{{\rm sim}}

\def\cc{{\rm cc}}

\titleformat{\section}{\normalfont\large\sc\centering}{\thesection}{1em}{}
\titleformat{\subsection}[runin]{\normalfont\large\bfseries}{\thesubsection}{1em}{}
\numberwithin{equation}{section} 
\renewenvironment{abstract}
               {\list{}{\rightmargin\leftmargin}%
                \item[\text{\hspace{10mm}\sc Abstract.}]\relax}
               {\endlist}

\begin{document}

\def\today{31-Jan-2026}
\def\heute{\today, one week after the Six-Minute race}

\begingroup
\begin{centering} 

\Large{\bf Six-Minute Man Sander Eitrem 5:58.52 -- \\
    first man below the 6:00.00 barrier}
    \\ [0.8em]
\large{\bf Nils Lid Hjort} \\[0.3em] 
\small {\sc Department of Mathematics, University of Oslo} \\[0.3em]
\small {\sc {\heute}}\par
\end{centering}
\endgroup


\begin{abstract}
\small{In Calgary, November 2005, Chad Hedrick was the
  first to skate the 5,000 m below 6:10. His world record
  time 6:09.68 was then beaten a week later, in Salt Lake City,
  by Sven Kramer's 6:08.78. Further top races and world records
  followed over the ensuing seasons; up to and including
  the 2024-2025 season, a total of 126 races have been below 6:10,
  with Nils van der Poel's 2021 world record being 6:01.56.
  The appropriately hyped-up canonical question
  for the friends and followers and aficionados of speedskating
  has then been {\it when} (and by whom)
  we for the first time would witness a below 6:00.00 race.
  In this note I first use extreme value statistics modelling
  to assess the state of affairs, as per the end of the 2024-2025
  season, with predictions and probabilities for the 2025-2026 season.
  Under natural modelling assumptions the probability
  of seeing a new world record during this new season
  is shown to be about ten percent. We were indeed 
  excited but in reality merely modestly surprised
  that a race better than van der Poel's record was clocked,
  by Timothy Loubineaud, in Salt Lake City,
  November 14, 2025. But Six-Minute Man Sander Eitrem's
  outstanding 5:58.52 in Inzell, on January 24, 2026,
  is truly beamonesquely shocking.
  I also use the modelling machinery to analyse the post-Eitrem
  situation, and suggest answers to the question of how fast
  the 5,000 m ever can be skated. 
  
\noindent
{\it Key words:}
extreme value statistics, 
five thousand metre speedskating,
Kuppern, 
models for records below thresholds, 
Sander Eitrem,
Sir Roger Bannister, 
Six-Minute Man 
}
\end{abstract}


\section{The 5k speedskating race, barrier by barrier}
\label{section:maxima}

\begin{figure}[h]
\centering
\includegraphics[scale=0.25]{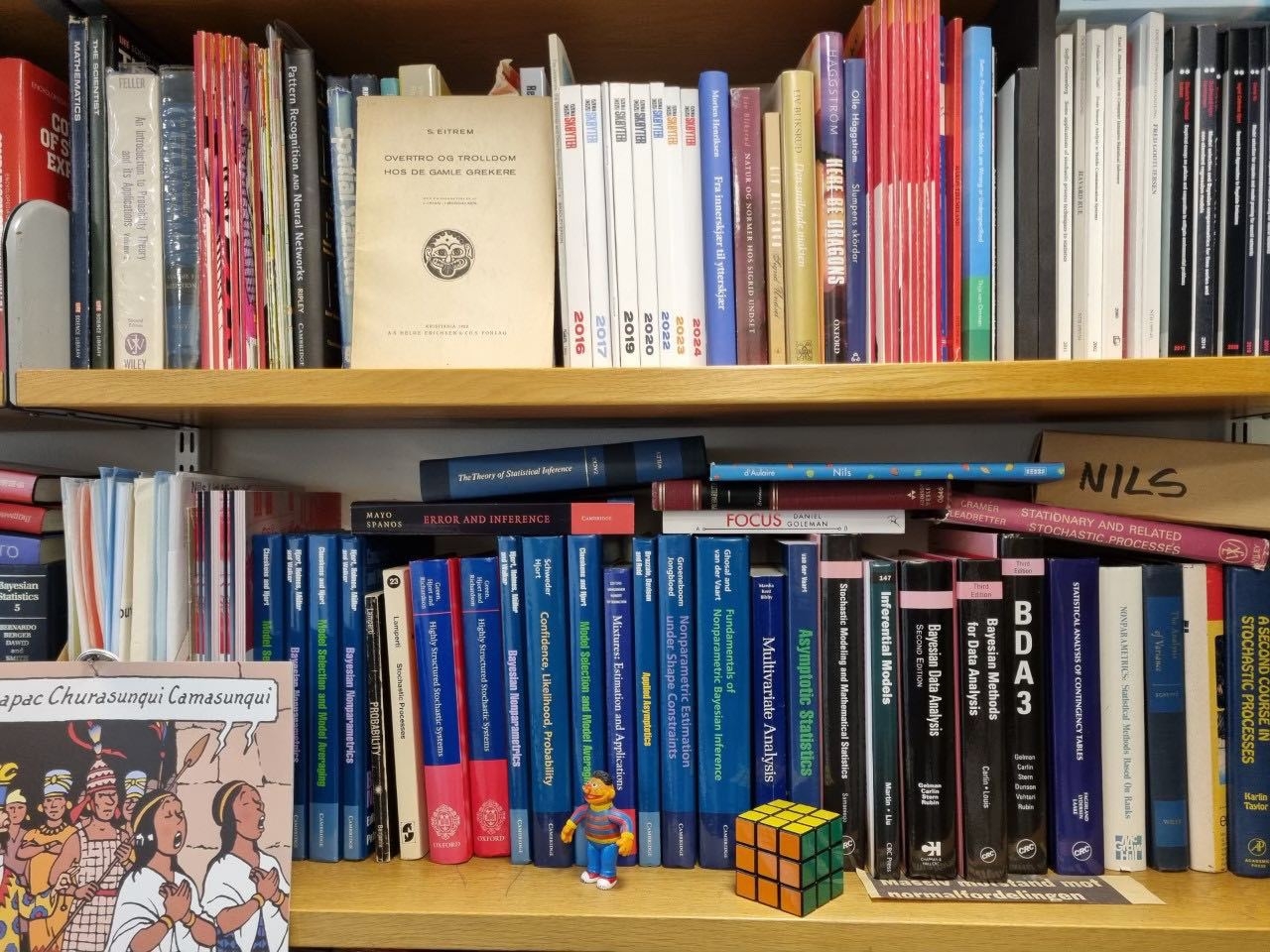}
\caption{
  There has of course been room for S.~Eitrem's
  {\it Overtro og trolldom hos de gramle grekere}
  on my bookshelves (from 1921, and with a new translation
  of Lukan's {\it L\o gnhalsen}), since S.~Eitrem first
  showed world-class 5k promise in 2021, i.e.~hundred years later.}
\label{figure:eitrem20}
\end{figure}

\noindent
The 5,000 m speedskating event is a classic one,
twelve and a half gruelling laps on the 400 m track
used for all serious competitions, since about 1895.
Skaters race in pairs (or, occasionally, in quartets,
though not for the major championships), currently
with speed up to unbelievable 50 km/h for the top skaters:
5 full kilometres, in just above 6 minutes!
This is more than twice as fast as the Jakob Ingebrigtsens
of the world manage in the absence of ice. 

In January 1955, Norwegian media were unwilling to believe it,
when first Dmitry Sakunenko (7:54.9) and then
Boris Shilkov (7:45.6, in a later pair) broke the 8-minute
barrier, in Medeo (in Kazakhstan, then CCCP);
even Norwegians were convinced a year later, however,
when Shilkov won the Olympic gold on the Lago di Misurina
skating rink with 7:48.7, eight seconds ahead of Sigge Ericsson.
In March 1965, Fred Anton Maier famously did 7:28.1
at Notodden (with Professor Svein Sj\o berg his pairmate,
setting a nice personal best of 7:58.5), the first below 7:30. 
As we recall, in sufficiently educated circles,
the next barrier was taken care of in Medeo March 1977,
first with Sergei Marchuk's 6:58.88 and then Kay Arne Stenshjemmet's
6:56.9 in a later pair.

Fast-feeding to the next barrier and the Nagano Olympics 1998,
an era later (and parenthetically bypassing fabulous skaters
Baranov, Pribytkov, Bochkarev, Shasherin, Visser, Karlstad, Gustafson,
Hadschieff, and even the Olympic quadruple-winner Johann Olav Koss,
all with fixed-blade skates, before the advantageous
clap-skates took over, in the season leading up to Nagano),
we have first Bart Veldkamp's 6:28.31 and then the gold medal
winner Gianni Romme's 6:22.22.

Since 1998, then, the world has been waiting for
the {\it six-minute man}. Progressive world records have been
set -- we recall Uytdehaage, Ervik, Hedrick (first under 6:10),
elegant Fabris 6:07.40 followed by an angry Kramer's 6:03.32 in 2007, 
Bloemen 6:01.86 in 2017,
van der Poel 6:01.56 in 2021;
as per the end of the 2024-2025 season, there
had been a total of $n=126$ races below 6:10. 

In this note I first study these fabulous sub-6:10 races,
showing that a two-parameter extreme value statistics model
fits these very well. This makes it possible to form
informative predictions for the 2025-2026 season,
including a clear probability curve 
\beqn
p(t)=\Pr({\rm there\ will\ be\ a\ race\ below\ }t),
\quad {\rm for\ }t\in[6\colon\!\! 00, 6\colon\!\! 10]. 
\eeqn
In particular, the meaningful $p({{\rm wr}})$ can be
computed, before the new season sets off, the probability
of seeing a race below van der Poel's 6:01.56. 
Importantly, the estimates of these $p(t)$ carry uncertainties,
of course, and these are strongly skewed. Even with
well estimated underlying parameters,
as with $\hatt p(t)=p(t,\hatt a,\hatt\sigma)$ below,
their distributions are skewed, leading to very
non-symmetric confidence intervals and curves, as
we shall see below. 

\section{A two-parameter model for sub-6:10 races}
\label{section:paramodel} 

\begin{figure}[h]
\centering
\includegraphics[scale=0.35]{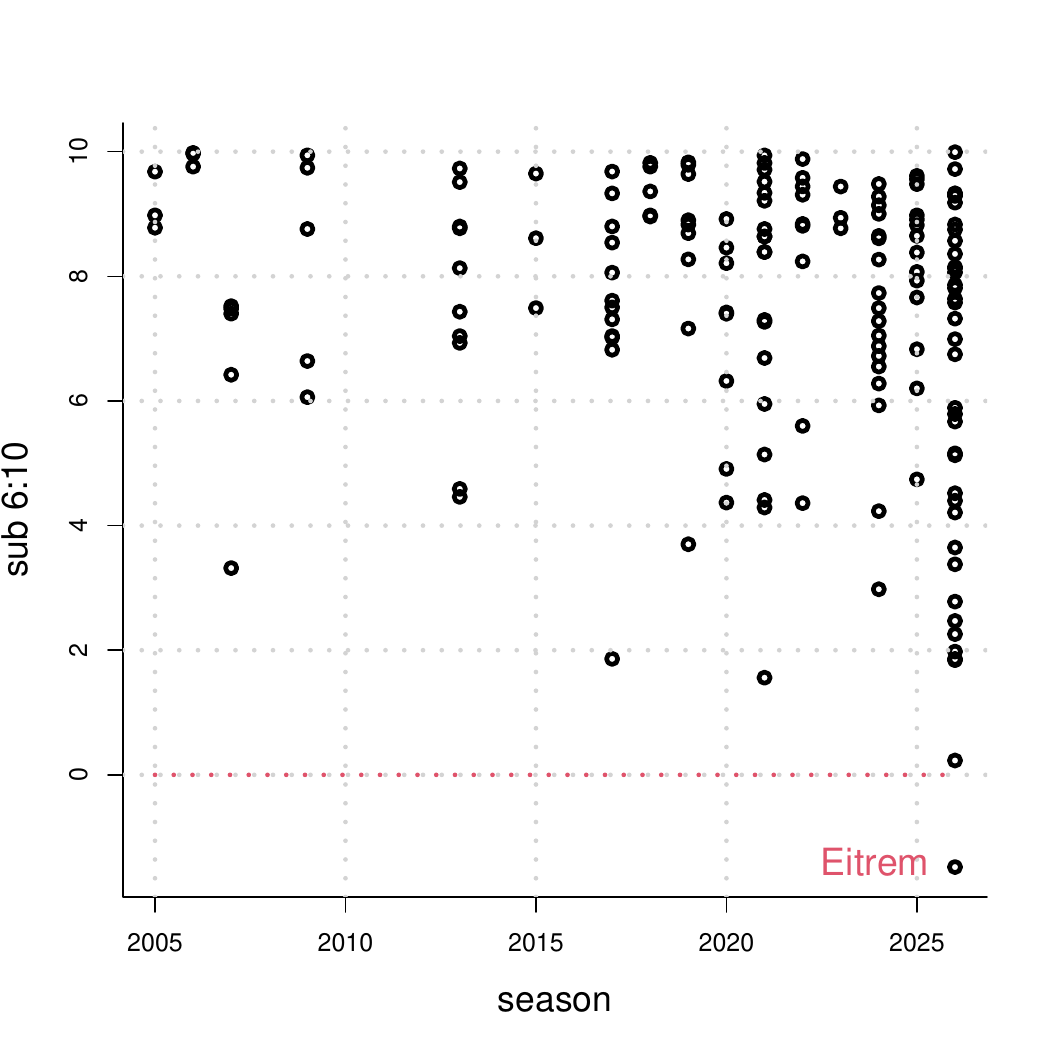}
\includegraphics[scale=0.35]{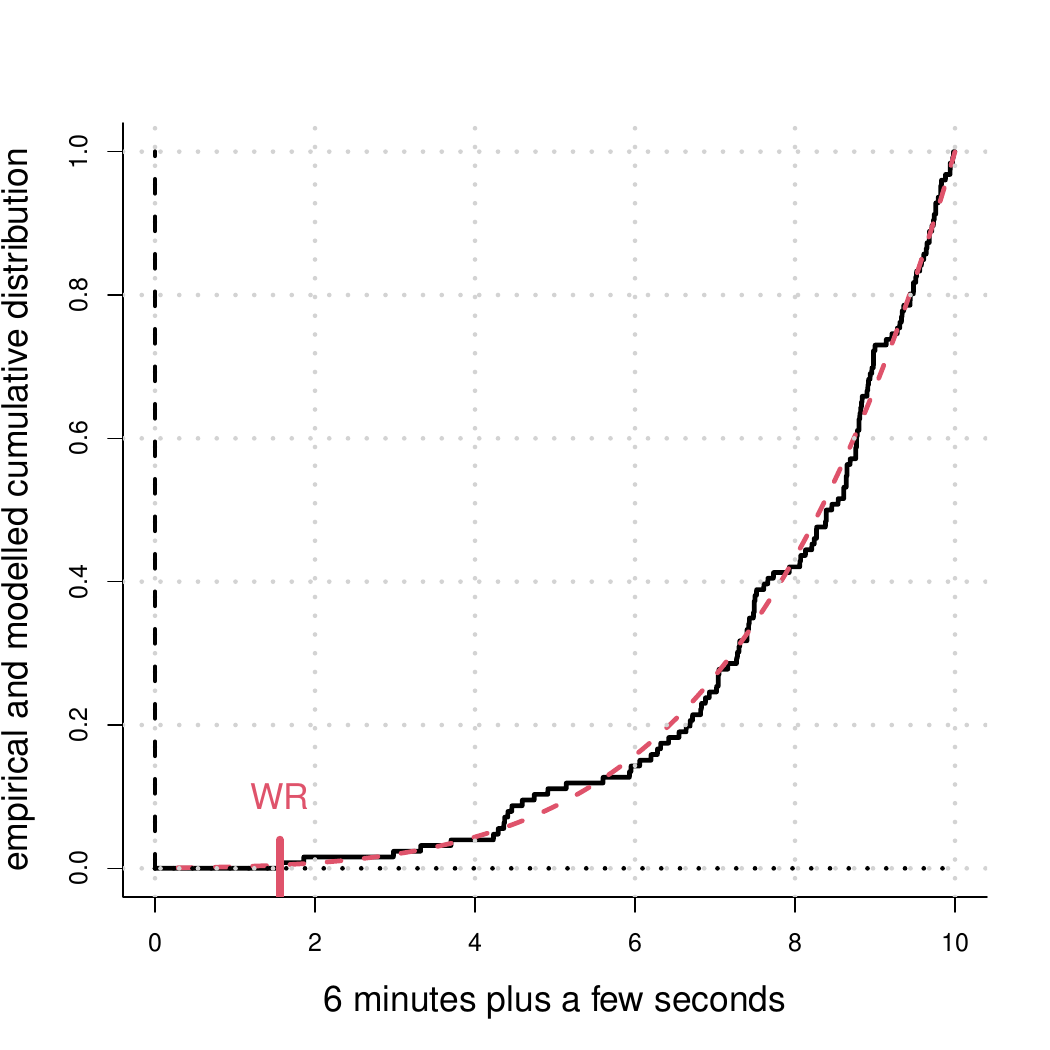}
\caption{
  Left panel: all 126 sub-6:10 races, from 2005 to 2025,
  where speedskating enthusiasts easily can spot and
  identify world records by Hedrick, Fabris, Kramer,
  Bloemen, van der Poel. Also included are the so far
  surprisingly high number 39 of such races for
  the 2025-2026 season, pre the Milano Olympics,
  with Eitrem's beamonesque 5:58.52.
  Right panel: empirical and fitted c.d.f.s 
  for sub-6:10 races, up to 2024-2025,
  with van der Poel's world record 6:01.56.}
\label{figure:fig21}
\end{figure}

The world's $n=126$ sub-6:10 races, up to the 2024-2025
season, are plotted in Figure \ref{figure:fig21}, left panel;
that figure also includes the so far 39 new sub-6:10 races
seen for the pre-Olympic 2025-2026 season. 
It is convenient here to translate race results $r_i$
to $y_i=6\colon\!\!10.00-r_i$, to more easily use established
extreme value statistics theory.
So van der Poel's world record corresponds to $y_{{\rm wr}}=8.44$,
and the top skaters have $y_i$ not far below this threshold.

The two-parameter model to be used here has cumulative and
density function
\beq
G(y,a,\sigma)=1-(1-ay/\sigma)^{1/a}
\quadandquad
g(y,a,\sigma)=(1-ay/\sigma)^{1/a-1}/\sigma. 
\label{eq:themodel}
\eeq
For such skating race data $a$ is most often positive,
so the allowed domain for $y$ are those for which
$ay/\sigma<1$, or $y\le \sigma/a$, a fixed upper bound,
corresponding to a fixed lower bound for the skating races $r$. 
The smooth limiting case $a\arr0$ is the exponential
$G(y,0,\sigma)=1-\exp(-y/\sigma)$; a slightly negative $a$
can be encountered too, with heavy tails and no fixed lower
bound for the world's top skaters. 

The log-likelihood function, for races $y_1,\ldots,y_n$,
which we take as being independent, is 
\beqn
\ell_n(a,\sigma)=\sumin \{-\log\sigma+(1/a-1)\log(1-ay_i/\sigma) \}. 
\eeqn 
For the $n=126$ sub-6:10 races, the maximum likelihood
estimates are $(\hatt a,\hatt\sigma)=(0.208,2.609)$,
with approximate standard errors $(0.083,0.314)$,
as per classical likelihood theory.
The empirical and fitted cumulative distribution
functions (c.d.f), i.e.~$G_n(y)$ and $G(y,\hatt a,\hatt\sigma)$,
translated back to race time scale, 
are graphed in Figure \ref{figure:fig21}, right panel;
as we see, the model fit is excellent, also when
it comes to world record terrain, the very best races. 

So via the $G(y,\hatt a,\hatt\sigma)$ we may predict
individual races,for skaters drawn from the pool of the
best ones. We are however interested in the best of
these races, for a certain event with e.g.~10 top skaters,
or over a full season. Via the machinery above,
we translate the best of $N$ races $r^*=\min\{r_1,\ldots,r_N\}$
to $Y^*=\max\{Y_1,\ldots,Y_N\}$. For this we have 
\beqn
\Pr(Y^*\le y_0)=G(y_0,a,\sigma)^N
   =\{1-(1-ay_0/\sigma)^{1/a}\}^N. 
\eeqn 
With $N$ a Poisson, with mean $\lambda$,
the as yet unknown number of races
below 6:10 in say the next season, we infer 
\beq
\Pr(Y^*\le y_0)=\exp\{ -\lambda (1-ay_0/\sigma)^{1/a} \}
   =p(y_0,a,\sigma,\lambda)
\label{eq:withpoisson} 
\eeq 
for the top result $r^*$ of such a season. 

\begin{figure}[h]
\centering
\includegraphics[scale=0.35]{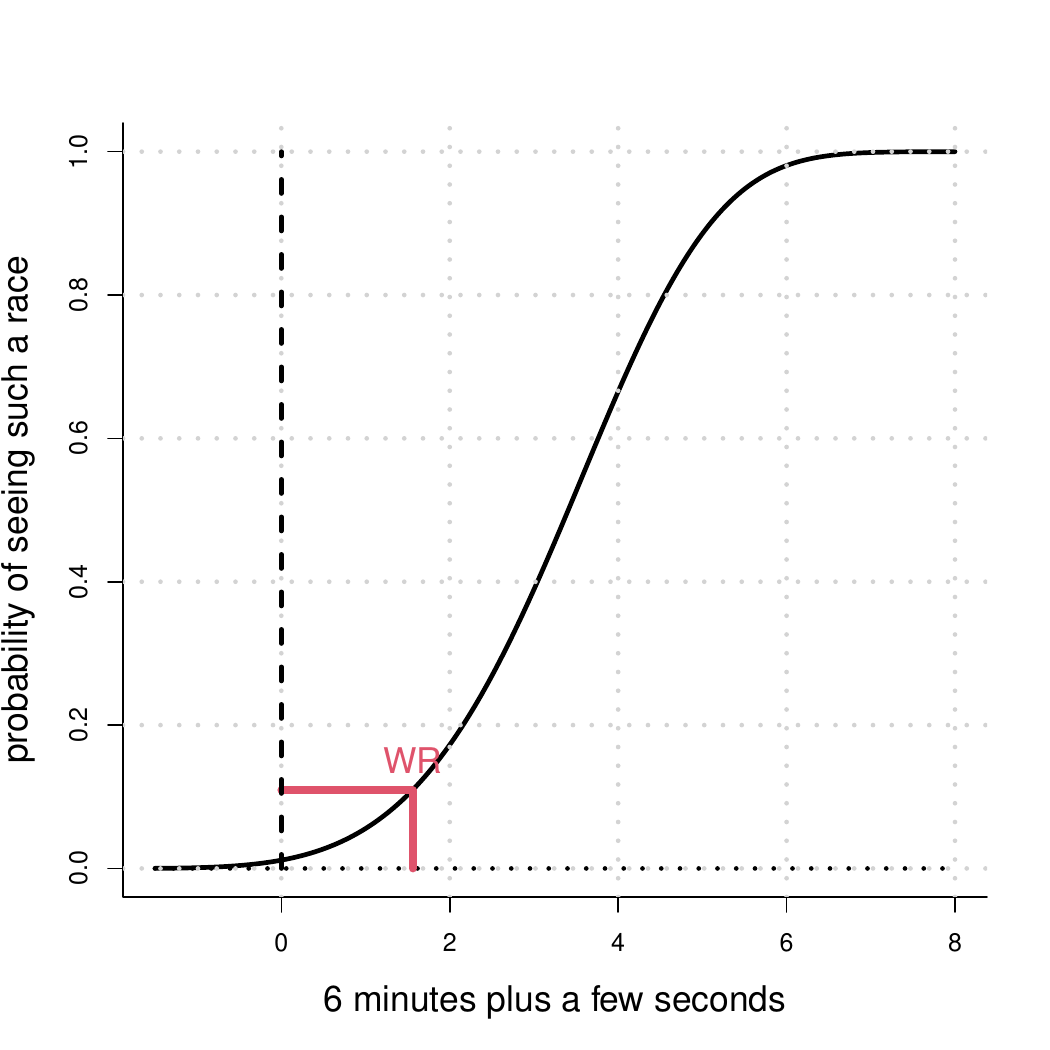}
\includegraphics[scale=0.35]{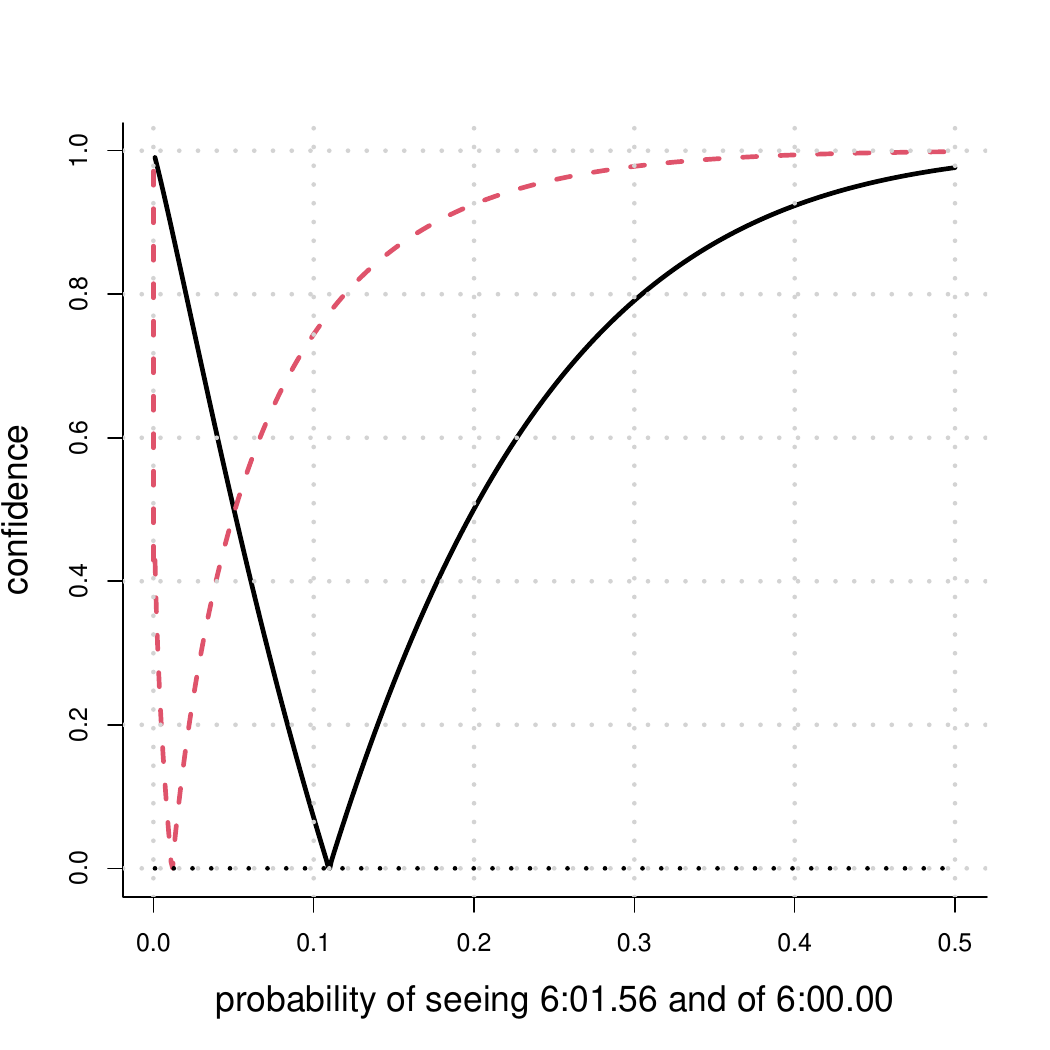}
\caption{
  Left panel: the probability curve, for seeing
  a race at a certain level, or better, during the 2025-2026 season,
  from the pre-season perspective,
  using $\lambda=25$ in the (\ref{eq:withpoisson}) formula.
  Ths probabilities for seeing a new world record,
  or a six-minute race, are estimated at 0.109 and 0.012. 
  Right panel: confidence curves for the two probabilities.
  These are non-tight and quite skewed, so quoting
  the 10.9 percent chance of seeing a world record
  is telling merely half the story.}
\label{figure:fig23}
\end{figure}

This may now be used to predict the best race of the 2025-2026
season, as seen from the pre-season perspective.
The number of sub-6:10 races over the past five
seasons have been 19, 9, 3, 18, 14, as per
Figure \ref{figure:fig21}, left panel. For an Olympic season
a higher number can be expected, with top efforts
not only from the main players among nations, but also from
others, so we may take e.g.~$\lambda=25$.
Applying (\ref{eq:withpoisson}) we find the 
estimated curve shown in Figure \ref{figure:fig23}, left panel,
of central interest to speedskating fans.
Those studying this curve, pre season, will see that there
is an approximate 10 percent chance of seeing a world record
during the 2025-2026 season; that seeing a best race
in the admirable time of 6:04 has about 50 percent chance;
and also that Six-Minute barrier just can be broken,
though with a low probability of about 1 percent. 

These probabilities are direct estimates, though, and
carry their uncertainties. The $p(y_0,\hatt a,\hatt\sigma,\lambda)$
quantities have quite skewed distributions, as it turns out,
so say 90 percent confidence intervals for the estimated
probabilities, here $0.109$ and $0.012$, will be skewed
and not symmetric. To compute appropriate confidence intervals
and curves, I follow methodology developed
in \citet[Ch.~3]{SchwederHjort16}, via Wilks theorems
and their refinements, with the main ingredient
being the log-likelihood profile function
\beqn
\ell_\prof(p)=\max\{\ell_n(a,\sigma)\colon p(y_0,a,\sigma,\lambda)=p\}. 
\eeqn 
Here $p$ is the estimate level, for any such probability
associated with Figure \ref{figure:fig23}, left panel.
This profile is then transformed to the confidence curve
\beqn
\cc(p)=\Gamma_1(D_n(p)),
\quad {\rm with\ the\ deviance\ }
D_n(p)=2\{\max\ell_\prof - \ell_\prof(p)\}, 
\eeqn 
with $\Gamma_1(\cdot)$ the c.d.f.~for the $\chi^2_1$. 

This confidence machinery is illustrated for
Figure \ref{figure:fig23}, right panel, for two
probabilities; (i) that of seeing a world record,
beating van der Poel's 6:01.56, estimated at 0.109;
and (ii) the Six-Minute probability of seeing a race
breaking that threshold, estimated at 0.012.
Confidence intervals are strongly skewed.
Thus merely saying, when a statistician might be
pressed to do so, that the new world record
probability is estimated at 0.109, is not quite
sufficient, as the uncertainty is significant;
a 90 percent confidence interval is $[0.006,0.440]$.
Similar comments apply, and more drastically,
to the Six-Minute estimate of 1.2 percent;
a 90 percent confidence interval associated
with that point estimate is $[0,0.233]$. 

\section{Testing model adequacy via monitoring processes}

\begin{figure}[h]
\centering
\includegraphics[scale=0.35]{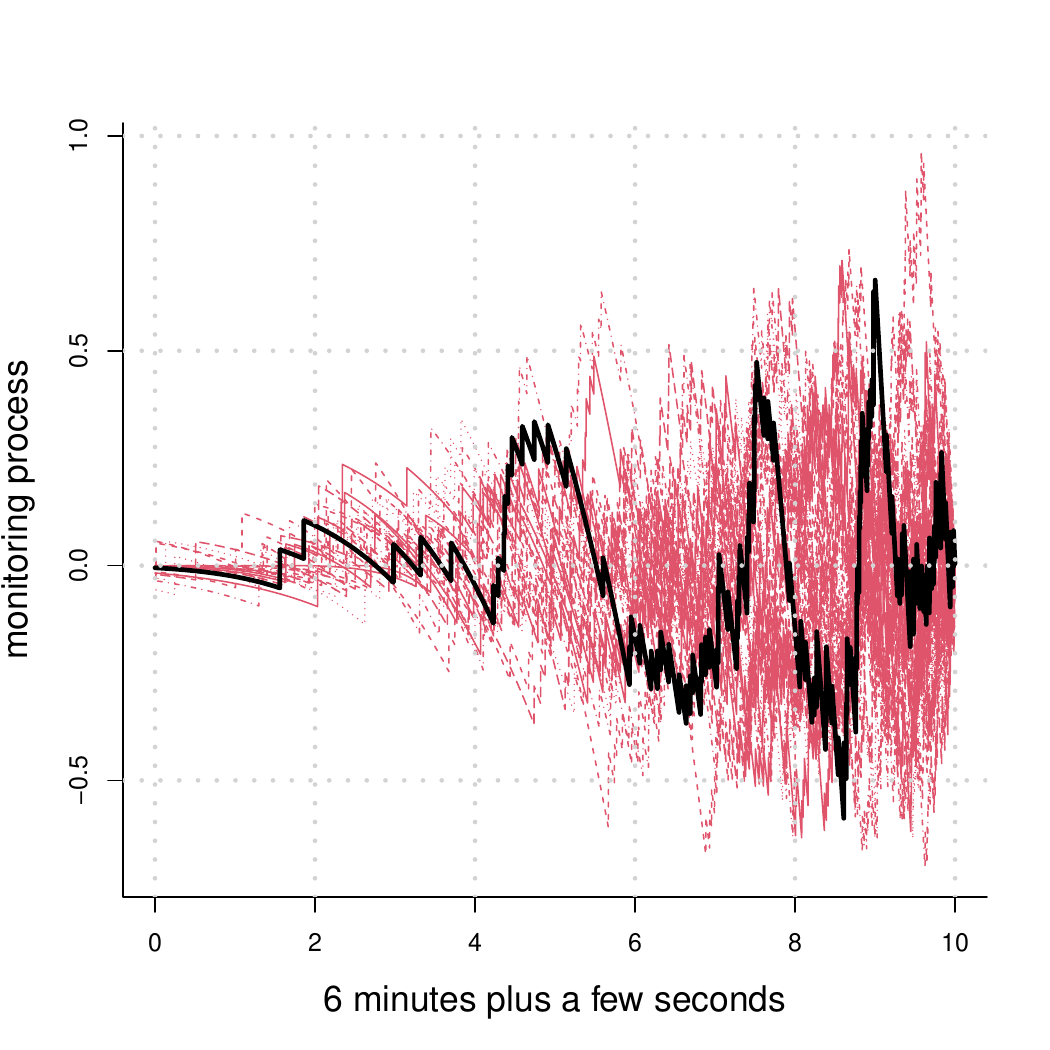}
\includegraphics[scale=0.35]{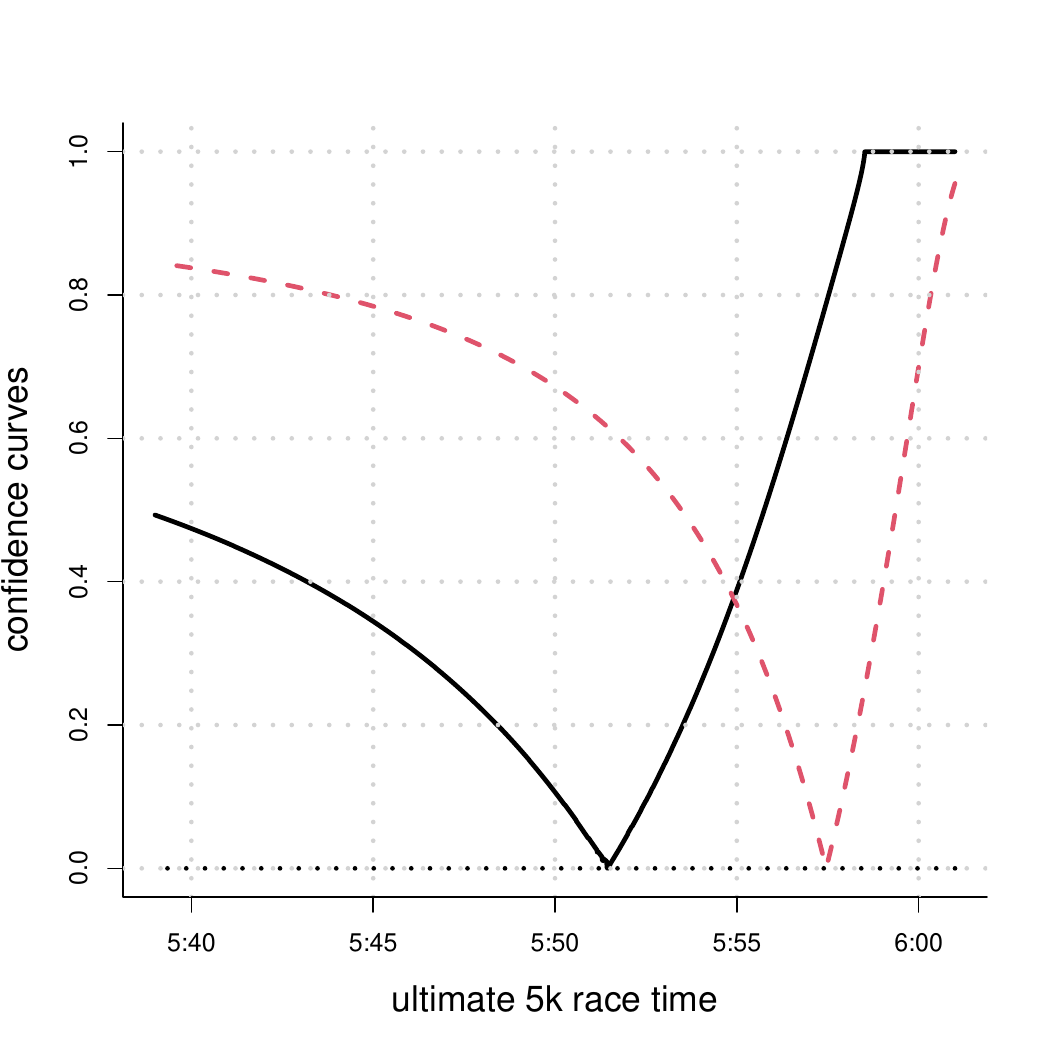}
\caption{
  Left panel: the model monitoring process $Z_n$
  based on all $n=126$ races from 2005-2006 to 2024-2025
  (full black line), along with $\simm=25$ simulated
  such processes (thin red lines), under model conditions.
  Right panel: the Ultimate Race time,
  $r_0=6\colon\!\!10.00-\hatt\gamma$, 
  estimated as 5:57.46 with pre-season data (red dashed) 
  but now post Eitrem as 5:51.26 (full black), 
  with confidence curves.}
\label{figure:fig26}
\end{figure}

The extreme value statistics machinery developed and applied
above have assumed that the 5k world is essentially
stable, or stationary, from the 2006-2006 to the 20024-2025
season. This statement concerns the level of the individual
top races, between 6:01.56 and 6:10.00, not the volume
of skaters managing to land in that admirable window.
To check this modelling assumption, consider the monitoring process
\beqn
Z_n(y)=\rootn\{G(y,\hatt a,\hatt\sigma)-G_n(y)\}
\quad {\rm for\ }y\in[0,10], 
\eeqn 
i.e.~the normalised difference between the parametric
and nonparametric cumulatives.
This is the full black curve inside the cloud
of similar but simulated thinner curves
of Figure \ref{figure:fig26}.

Large-sample theory can be be developed
for the $Z_n(y)$ process, and it converges indeed under model
conditions in distribution to a well-defined zero-mean
Gaussian process; see \citet[Ch.~10]{HjortStoltenberg26}
for technical details. For the present purposes it is
easier to simulate $Z_n$ processes, however, under model
conditions, to see how the observed $Z_n$ sits inside
such a cloud. These are the $\simm=25$ thin curves
in the left hand panel. For each, there is a simulated
dataset $y^*=(y_1^*,\ldots,y_n^*)$, constructed via
\beqn
y_i^* = (\hatt\sigma/\hatt a)\{1-(1-u_i)^{\hatt a}\} 
\quad {\rm for\ }i=1,\ldots,n, 
\eeqn 
with the $u_i$ an i.i.d.~uniform sample. 
We learn from this that the observed monitoring process
is fully inside the normal range for such curves,
i.e.~there is no statistical evidence against
the stationarity assumption from 2005-2006 to 2024-2025. 

Another check of interest, regarding the stationarity,
is to see if the $(a,\sigma)$ parameters have remained
essentially constant, over this nineteen-year window.
A model which should be able to pick up any trend,
if present, takes 
\beqn
(a_j,\sigma_j)=(a,\sigma\exp(\gamma x_j))
\quad {\rm for\ }j=1,\ldots,19, 
\eeqn 
with $x_j=t_j-\bar t$, say, calendar year minus their average.
Fitting and assessing this model shows that the estimated $\hatt\gamma$
is rather close to zero, i.e.~no significant trend
is present, inside this period.

Crucially and revealingly, however, when the dataset is
extended, from the first ever 2005-2006 season with any sub-6:10
up to and including the first half of the Olympic 2025-2026
season, there {\it are} changes, {\it both} in volume
(39 new such top races, during November, December, January)
{\it and} in quality. This matches what can be interpreted
from the raw data, plotted in Figure \ref{figure:fig21}, left panel.
New skaters in the very top category,
like Loubineaud, \jilek, Dawson, Semirunniy, Lorello,
Fartsholder (as Klassekampen rather aptly writes his name,
see \citet{Hjort25B}), along with established giants
like Ghiotto, Bloemen, Bergsma, Huizinga, form a significantly
faster 5k ensemble, compared to the top skaters of the previous seasons. 

\section{How fast can they race?} 
\label{section:hjallis}

``Du kainn itj g\aa{} p\aa{} null'', opined Hjallis,
a.k.a.~triple-gold Olympic winner Hjalmar Andersen, 
when asked about how fast one could ever skate the 10,000 m.
For the extreme values model (\ref{eq:themodel})
we may at least estimate that alluring quantity,
a lower bound $r_0$ for how fast skaters can manage the 5k.
When $a$ in the model is positive, which for sports results
is a natural assumption, the upper limit is the clear-cut parameter
\beqn
\gamma=\sup\{y>0\colon g(y,a,\sigma)>0\}=\sigma/a,
\quad {\rm translating\ to\ }r_0=6\colon\!\!10.00-\gamma. 
\eeqn
When $a$ tends to be small, the $\gamma$ is high
and $r_0$ is small; in the smooth limit, where $a$ tends
to zero, the $Y$ distribution is the exponential,
with infinite support, which here would correspond to
there being no clear lower limit for speedskating races. 

The maximum likelihood estimate may be read off
from earlier estimation efforts as $\hatt\gamma=\hatt\sigma/\hatt a$.
The traditional delta method ensures that approximate
variances for smooth functions of $(\hatt a,\hatt\sigma)$
can be read off via quadratic forms of the inverse
ovserbed Fisher information matrix, Here we are
outside that smooth terrain, due to $a$ being small,
so more precise analysis comes via the log-likelihood profile
function
\beqn
\ell_\prof(\gamma)=\max\{\ell(a,\sigma)\colon \sigma/a=\gamma\}
   =\max_{{\rm all\ }a} \ell(a,a\gamma). 
\eeqn 
Again following the methods developed in \citet[Ch.~3]{SchwederHjort16},
the confidence curve for $\gamma$, which then translates
to such for the ultimate race time $r_0$, is
\beqn
\cc(\gamma)=\Gamma_1(D_n(\gamma)),
\quad {\rm with\ deviance\ }
D_n(\gamma)=2\{\max\ell_\prof - \ell_\prof(\gamma)\}. 
\eeqn 

I have carried out such analysis for the case of
pre 2025-2026 data, when the top 5k world was reasonably
stationary, with world record 6:01.56, and then
for the revised world view after Eitrem and the Inzell
24-Jan-2026 World Cup races.
Pre-season, we have $\hatt\gamma=12.54$,
with $\hatt r_0=5\colon\!\!57.46$; with the fabulous races
for the 2025-2026 season, I find $\hatt\gamma=18.74$,
with $\hatt r_0=5\colon\!\!51.26$.
See Figure \ref{figure:fig26}, right panel,
with confidence curves. These are and have to be
quite wide, as there are inherent uncertainties
in the very lower tail of the top races.


Eitrem's race had a 200 m opening of 19.03,
followed by twelve stunning 400 m laps averaging 28.39
seconds for each. The presumed and estimated Ultima Thule 5k Race
would mean and openiing of 18.00 and then twelve
close to unbelievable laps with 27.77;
see \citet{Lid1942}. 
We shall see -- and we all remember Yuskov's race
in Salt Lake City in November 2013, for its first seven laps,
and we know from Semerikov and Dawson how the last five laps
of a 5,000 m occasionally can end up looking like 1500 m laps.
We will excitedly and fully believe it, when Stolz does the 1500 m
in 1:40 (perhaps next year), or 100 seconds, which is 54.0 km/h speed,
but it is currently beyond human belief systems that
he could go on for a full 5k in that speed
(which would mean an another-planetary 5:33.33). 




\section{Concluding remarks} 
\label{section:concluding}

In this note I have examined the full set of 126 sub-6:10
races, from Hedrick's world record November 2005 up to the end
of the 2024-2025 season, fitted these to the well-fitting
two-parameter model (\ref{eq:themodel}), and then used
the appropriate machinery to predict future best-race results,
for a given event, like a World Cup, or over a full season. 
Some rounding off comments are as follows. 

\smallskip
{\it Remark A: Eitrem.}
While we await a high quality post-career biography
to be written by or about Eitrem, ten years from now
(perhaps by Jon Fosse or Anders Heger), 
there is ample material to be found in skating literature
and on the net. 
A detailed and redable account on his career, from
young child to World Champion and now world record holder,
is in \citet[pages 6-10]{Finsen26};
his {\it Alt om sk\o yter 2026} indeed serendipitously
has Eitrem as front cover man. 

\smallskip
{\it Remark B: volume, quantity, horizon.}
I have indicated above that the 2025-2026 season
is indeed a significant break with the stationarity
of 2005-2006 to 2024-2025, in both volume and quality.
Remarkably, the international spread is also
widening, outside the traditionally strongest nations.
The full list of sub-6:10 nations, represented by
their usual acronyms, is
NOR, FRA, SWE, USA,
CAN, CZE, NED, AUT,
ITA, GER, KOR, POL,
CHN, RUS, BEL, JPN, NZL. 
Check indeed Section \ref{section:nationalrecords},
with items most enthusiasts know by heart. 

\smallskip
{\it Remark C: Bannister and Eitrem.} 
Eitrem's 5:58.52 is for the history books and
for the eternal memory of speedskating. Even {\it Dagsrevyen}
at the NRK, the canonical main channel for the Norwegian
Broadcasting Corporation, and which very unfortunately
shies away from mentioning speedskating at
all,\footnote{thanks to the intricacies of modern
attention economy and competition with other television channels; 
enthusiasts are pushed to pay NOK 399 pro month
to watch our heroes Eitrem, \jilek, Loubineaud, Ghiotto,
Bergsma, Kongshaug; see \citet{Hjort25B} for further
lamentations} spent a national record time of nine minutes on him.

In the history of sports, media, culture,
communication, exposure, public attention,
nothing can beat Sir Roger Bannister, however;
read his {\it The First Four Minutes} \citep{Bannister55}.
This had to do both with
the fabulous race itself, at the Iffley Road track
in Oxford, May 1954, but also the media situation
ten years after the war, 
its immediately understood highly symbolic value,
the receptibility of both UK and the rest of the world.
This was not merely a nerd story for those sufficiently
interested, but world news, for every citizen
and family, for politics and dinner tables.
I fear, though, that there are still segments
of people not yet knowing about, or understanding,
or appreciating, the Six-Minute Man. But we should muster
the courage to say, even if we do not dare to shout it from
our modest rooftops, that Sander Eitrem is Norway's answer
to Sir Roger Bannister. It is indeed directly comparable
to Kuppern's 15.46.6 in Squaw Valley 1960 (Kosichkin
skated in a later pair), a feat discussed or pointed to
in at least four novels by prominent Norwegian authors. 

Sir Roger's Four-Minute Mile also had a certain 
`ketchup effect'; once {\it he} was under four, other
athletes understood they could do it too. His record
was broken the same year, by John Landy, and within
two years about ten runners had achieved the sub-four.
So perhaps also Eitrem's feat will be followed by ketchup. 
But it may also prove to be a Jim Hines phenomenon;
as we recall he was the first to sub-Hary with his 9.95
in the thin-air Mexico City Olympics 1968, and 14 years
had to pass before we saw 9.93. 

\smallskip
{\it Remark D.}
My model (\ref{eq:themodel}) is a special case
of more general model formulations for extreme value
statistics, built up during several decades,
associated with the names of Fisher, Kolmogorov, 
Gnedenko, Tippett, Gumbel, and several others.
Interestingly, both E.~Gumbel and S.~Eitrem
wrote about lies and liars, around 1921, a century ago;
check with \citet{Eitrem1921}, \citet{Gumbel19, Gumbel22, Gumbel58}.


\smallskip
{\it Remark E: why and how are records beaten?}
Suppose $X_1,X_2,\ldots$ are i.i.d.~from some continuous
distribution, with $M_n=\max_{i\le n}X_i$ the best result
after $n$ trials; this is a new record provided $M_n>M_{n-1}$.
Crucially, such records are few and far between,
under the i.i.d.~assumption. With $Z_n$ the number
of records set, in the course of the first $n$ trials,
$\E\,Z_n$ is the harmonic sum $1+1/2+\cdots+1/n$, growing
slowly, as $\log n$. Indeed it can be shown, via Lindeberg theorems,
that $(Z_n-\log n)/(\log n)^{1/2}\arr_d\N(0,1)$.
So why and how are many more records set, after all?
The answer is that the i.i.d.~condition does not hold up;
equipment, ice conditions, halls, training regimes become steadily
better. I trust, though, that 5:59.99 will still be seen
as a pretty good result, twelve years from now. 

\smallskip
{\it Remark F: other barriers to break.}
Methods developed and applied in this note may be used
for similar studies, with probability assessments
and predictions, of the other speedskating distances,
to various other sports, and to various other
phenomena involving measurements setting new records.
Methods similar to those exhibited here
are indeed used to assess Usain Bolt's world records
in the 100 m sprint, in \citet[Ch.~14]{SchwederHjort16},
for instance, and some applications to temperature
time series in meteorology are briefly outlined
in \citet{Hjort26}. See also \citet{Hjort26B} for
studying chess ratings, though with different
parametric models.
Magnus Carlsen is the Sander Eitrem of chess. 

Vital speedskating thresholds ahead of us include
the following (where we ought to check with Jordan Stolz
in a few years, regarding the first three). 
(i) The first 33.50 or better in the 500 m;
(ii) the first 1:05.00, or 65 seconds, in the 1000 m; 
(iii) the first below 1:40.00, or 100 seconds, in the 1500 m;
(iv) the first lady below 6:35 in the 5k;
(v) the first lady below 15:30 in the 10k.

I have actually started to look at the 500 m,
starting with all races, so far, below 34.50 seconds.
Here there is a detectable trend of improvement,
over recent years, however, unlike the 5,000 m
over the 2005-2006 to 2024-2025 period,
where stationarity was seen to be an adequate
assumption in the analysis above. This means that 
the model (\ref{eq:themodel}) needs adjustment,
with the parameters $(a,\sigma)$ developing over time.
Analysis and predictions hence become more complex,
but are doable with extended care and efforts. 

\smallskip
{\it Remark G: quartets.}
Quartets are for Beethoven and Shostakovich, and, occasionally,
for 5k speedskating. The world would have been
less satisfying if the six-minute had been broken
that way, however; quartet skaters have a slight time advantage. 




\section*{Acknowledgments} 

Speedskating historians are always indebted to the
admirable efforts of Evert Stenlund
and his website, {\tt www.evertstenlund.se/other.htm},
along with {\tt speedskatingresults.com}
and {\tt live.isuresults.eu/events}. 
These online databases and tables are wondrously
and consistently useful and continuously updated.
The data I have used in this report are essentially 
from the Stenlund site,
patiently excerpted, reorganised, curated,
supplemented with my own personal notes over the past seventy years
or so. Thanks are also due to the many active and knowledgeable
people in the {\it Forum for sk\o ytehistorie} on Facebook,
where I serve among the three admins.

\section*{National records in the sub-6:10 club, as of \today}
\label{section:nationalrecords} 

\begin{footnotesize}
\begin{verbatim}
  5.58.52 EITREM Sander            NOR Inzell      26 0124  1       
  6.00.23 LOUBINEAUD Timothy       FRA SLC         25 1114  1       
  6.01.56 van der POEL Nils        SWE SLC         21 1203  1        
  6.01.84 DAWSON Casey             USA Calgary     25 1121  1     
  6.01.86 BLOEMEN Ted-Jan          CAN SLC         17 1210  1        
  6.01.98 JILEK Metodej            CZE Inzell      26 0124  2   
  6.02.98 ROEST Patrick            NED SLC         24 0128  1        
  6.04.21 FARTHOFER Alexander      AUT SLC         25 1114  1       
  6.04.23 GHIOTTO Davide           ITA SLC         24 0128  2
  6.06.75 GROSS Gabriel            GER SLC         25 1114  3      
  6.07.04 LEE Seung-Hoon           KOR Calgary     13 1110  3
  6.07.81 SEMIRUNNIY Vladimir      POL Inzell      26 0124  6     
  6.08.36 LIU Hanbin               CHN Calgary     25 1121  4       
  6.08.64 TROFIMOV Sergej          RUS SLC         21 1203  1
  6.08.76 SWINGS Bart              BEL SLC         21 1203  4
  6.08.83 SASAKI Shomu             JPN SLC         25 1114  4       
  6.09.68 MICHAEL Peter            NZL SLC         17 1210  7
\end{verbatim} 
\end{footnotesize}

\noindent 
We're eagerly waiting for SUI, FIN, DEN, KAZ, BLR, EST, LAT
and yet other civilised societies to join this list.

\begin{small}

  
\bibliographystyle{biometrika}
\bibliography{diverse_bibliography2026}

@book{bannister55,
author = {Bannister, R.}, 
title = {The First Four Minutes},
publisher = {Putnam}, 
address = {London}, 
year = {1955},
}

@book{Eitrem1921,
author = {Eitrem, S.},
title = {Overtro og trolldom hos de gramle grekere}, 
publisher = {A.S. Helge Erichsens forlag}, 
address = {Kristiania},
year = {1921},
}

@book{finsen26,
author = {Finsen, L.}, 
title = {Alt om sk\o yter 2026}, 
publisher = {Ortygia}, 
address = {Redal, Naustdal},
year = {2026},
}

@book{Gumbel19,
author = {Gumbel, E. J.},
title = {Vier Jahre L\"uge},
publisher = {Verlag Neues Vaterland},
address = {Berlin},
year = {1919},
}

@book{Gumbel22,
author = {Gumbel, E. J.},
title = {Vier Jahre Politischer Mord},
publisher = {Verlag der neuen Gesellschaft},
address = {Berlin},
year = {1922},
}

@book{Gumbel58,
author = {Gumbel, E. J.},
title = {Statistics of Extremes},
publisher = {Columbia University Press},
address = {Cambridge},
year = {1958},
}

@article{Hjort25B,
author = {Hjort, N. L.},
title = {Sk\o ytel\o penes uutholdelige letthet},
journal = {Klassekampen.}, 
note = {Kronikk, Maier day 15-xii-2025}, 
month = {dec}, 
year = {2025},
}

@article{Hjort26B,
author = {Hjort, N. L.},
title = {Anyone for chess? {A}nalysing chess ratings
   above high thresholds}, 
journal = {FocuStat Notes},
volume = {12},
pages = {1--8},
year = {2026},
}

@book{HjortStoltenberg26,
author = {Hjort, N. L. and Stoltenberg, E. {Aa}.},
title = {Statistical Inference: 600 Exercises and 100 Stories},
publisher = {Cambridge University Press},
address = {Cambridge}, 
year = {2026},
}

@incollection{Hjort26, 
author = {Hjort, N. L.}, 
title = {The stochastic view used in climate sciences:
(some) perspectives from (some of) mathematical statistics}, 
booktitle = {The Hasselmann Legacy:
  Stochastic Thinking in Climate Science}, 
editor = {Lin Lin and Hans von Storch},
publisher = {Springer Verlag}, 
address = {Berlin}, 
pages = {xx--xx}, 
year = {2026}, 
}

@article{Lid1942,
author = {Lid, Nils}, 
title = {Ultima {T}hule},
journal = {Trolldom, Nordiske studier},
volume={i}, 
pages = {165--179}, 
year = {1942},
}

@book{SchwederHjort16,
author = {Schweder, T. and Hjort, N. L.},
title = {Confidence, Likelihood, Probability: 
   Statistical Inference with Confidence Distributions},
publisher = {Cambridge University Press},
address = {Cambridge},
year = {2016},
}

\end{small}

\begin{small}

\section*{Footnote} 

After having been appropriately absorbed by the World
Cup events and various championships, during the first
months of the 2025-2026 season, 
via speedskating fora and the unfortunate NOK 399 pro month
to get Viaplay Vinter on my laptop screen, 
I was inspired to dig out data and write up this note.
I then wished to write up a good {\it FocuStat Blog Post},
as these are sometimes reaching a sizeable international
readership (we've got fanmail from Steven Pinker, etc.).
To my frustration {\it\&} irritation the University of Oslo
has changed the Wondrously Well-Working Vortex {\tt html},
which I and FocuStat comrades have used for ten years,
for a Horribly Bad-Working Vortex editing system, where
writing math is a nightmare; having the new Vortex system
installed has also seriously disturbed some of our previously
perfect pitch layouted Blog Posts. I might have to move
to Andorra to find a university taking better care
of its humble employees. 

\end{small}



\end{document}